\documentclass[12pt,a4paper]{article}
\usepackage{amsmath}
\usepackage{amssymb}
\usepackage{color}
\usepackage{graphicx}
\usepackage{float}
\usepackage{cite}
\usepackage{dcolumn} %% tables cols aligned at decimal point
\newcolumntype{.}{D{.}{.}{1}}
\usepackage[ruled, vlined, linesnumbered]{algorithm2e}
\usepackage{setspace}

\usepackage{algpseudocode}              % for writing pseudocode easily
\usepackage{mathrsfs}
\usepackage{booktabs,threeparttable}
\usepackage{mhchem}
\usepackage{braket}
\usepackage{dsfont}

\usepackage{paralist}
\usepackage{xcolor}
\usepackage[normalem]{ulem}

\begin{document}
\begin{center}

\vspace*{0.5cm}

{\Large\bf
Gaussian Process Regression Adaptive Density-Guided Approach:
Towards Calculations of Potential Energy Surfaces for Larger Molecules 
}

\vspace{1.5cm}

{\large Denis G. Artiukhin$^{\parallel}$\footnote{Email: denis.artiukhin@fu-berlin.de}, 
Ian H. Godtliebsen$^{\dagger}$\footnote{Email: ian@chem.au.dk},
Gunnar Schmitz$^{\mathparagraph}$\footnote{Email: gunnar.schmitz@rub.de}, \\
and Ove Christiansen$^{\dagger}$\footnote{Email: ove@chem.au.dk} \\[2ex]
}

\vspace{1.5cm}
$^{\parallel}$
Institut f\"ur Chemie und Biochemie, Freie Universit\"at Berlin, \\ Arnimallee 22, 14195 Berlin, Germany \\[1ex]
\vspace{0.5cm}
$^{\dagger}$
Department of Chemistry, Aarhus Universitet, DK-8000 Aarhus, Denmark \\[1ex]
\vspace{0.5cm}
$^{\mathparagraph}$
Lehrstuhl f\"{u}r Theoretische Chemie II, Ruhr-Universit\"at Bochum, \\ Universit\"{a}tstraße 150, 44801 Bochum, Germany\\[1ex]
\vspace{0.5cm}

\end{center}

\vfill

\begin{tabbing}
Date:   \quad\= March 27, 2023 \\
%Status: \> to be submitted\\
\end{tabbing}.

\newpage

\begin{abstract}

We present a new program implementation of the gaussian process regression adaptive density-guided approach 
[\textit{J.\ Chem.\ Phys.} 153 (\textbf{2020}) 064105]
in the {\sc MidasCpp} program. A number of technical and methodological improvements 
made allowed us to extend this approach towards calculations of larger molecular systems than those 
accessible previously and maintain the very high accuracy of constructed potential energy surfaces.
We demonstrate the performance of this method on a test set of molecules of growing size and show that  
up to 80~\% of single point calculations could be avoided introducing a  
root mean square deviation in fundamental excitations of about 3~cm$^{-1}$.
A much higher accuracy with errors below 1~cm$^{-1}$ could be achieved with tighter convergence thresholds
still reducing the number of single point computations by up to 68~\%. 
We further support our findings with a detailed analysis of wall times 
measured while employing different electronic structure methods.
Our results demonstrate that GPR-ADGA is an effective tool, which could be applied for cost-efficient 
calculations of potential energy surfaces suitable for  
highly-accurate vibrational spectra simulations.

\end{abstract}

\newpage
\clearpage

\section{Introduction \label{sec:intro}} 

Constructions of potential energy surfaces (PESs) are of vast interest for many fields of chemistry as they can provide 
a detailed insight into dynamics and reactivity of molecules. 
Although the dimensionality of the PES increases linearly with the number of nuclei $K_{\mathrm{nuc}}$, i.e.,  
it is equal to the number of vibrational modes $M = 3 K_{\mathrm{nuc}} - 6(5)$, the computational cost of the PES construction 
scales exponentially due to the need to compute mode--mode coupling terms. 
As the result, calculations of accurate fully-coupled PESs are prohibitively expensive 
and are only possible for up to four-atomic molecules. 
A commonly used approach to reduce the computational cost of the PES construction 
relies on the restriction of high-order mode couplings
and is known under several names such as the $n$-mode expansion~\cite{jung1996,carter1997,bowman2003,rauhut2004,kongsted2006}, 
cluster expansion~\cite{meyer2012}, or high-dimensional model representation~\cite{rabitz1999}.
This method reduces the cost of the PES generation significantly and enables calculations of up to a few dozens of atoms.
A large number of different strategies could be applied to further decrease the cost of $n$-mode-expanded PESs.
These include, to name but a few, 
many-body expansions~\cite{koenig2016,madsen2018},  
approximate computations of high-order coupling terms~\cite{rauhut2004,yagi2007,rauhut2008,Rauhut2009,sparta2009_2,sparta2010,meier2013,schmitz2019},
various screening techniques~\cite{benoit2004,rauhut2004,benoit2006,pele2008,benoit2008,seidler2009,cheng2014,Klinting2020}, 
vibrational space dimensionality reduction~\cite{benoit2004,mackeprang2015,yagi2019},
and the use of molecular symmetry~\cite{sparta2010,ziegler2018}.

If PESs are constructed on a grid, the corresponding computational cost could be high due to
a non-optimal spacial placement of individual grid points and, as the result, a large number of them. 
The design of good ways of constructing the grid of points, to avoid a too
large computational burden, can requiry significant human time. 
The adaptive density-guided approach 
(ADGA)~\cite{sparta2009,toffoli2011,klinting2018} is designed to mitigate these issue. 
It constructs grids within the $n$-mode representation in an iterative procedure being guided by one-mode vibrational densities.
The ADGA ensures that a modest number of single points (SPs) is computed 
while maintaining a very high accuracy of the PES. Furthermore, it enables a fully automatic determination of the grid dimensions and granularity
without using any prior knowledge about the molecular system. Having proved to be a reliable method for PESs computations, 
the ADGA was further extended methodologically with various algorithms for the grid boundary extension~\cite{klinting2018},
the use of energy derivatives and molecular point groups of symmetry~\cite{sparta2010},
and different fitting functions for the analytical representation of PESs~\cite{klinting2018}.
Additionally, combinations of the ADGA with multiresolution PESs computations~\cite{sparta2009_2} and double incremental PESs expansions~\cite{arti2020}
were presented.

The use of machine learning (ML) algorithms for constructing PESs and/or assisting in their computations is 
also gaining its momentum. Thus, neural networks (NNs) were already successfully
employed for PES representation and for vibrational structure 
calculations~\cite{Manzhos2006, Manzhos2008,Brown2017,Pradhan2017,BrownHFCO,Pradhan2016}.
Furthermore, in case of molecular dynamic simulations, it was demonstrated that NNs can 
extend the simulation time and treat large molecular systems with an accuracy similar to that of 
density functional theory~\cite{behler2007,behler2011,behler2017,Behler2016,Ko2021}.
Gaussian Process Regression (GPR)~\cite{rasmussen2005}, a nonparametric Bayesian ML approach, 
deserves special attention as it provides uncertainties for predicted data points.
This allows to estimate the quality of the fit at regions of interest and 
make decisions on whether these regions should be supplied with additional training data points.
Note, however, that uncertainty estimates could also be made by combining several NNs in a committee~\cite{Schran2020}. 
This characteristic of GPR makes it well-suited for the use in Bayesian optimization and active learning.
In this regard, Jinnouchi et al.~\cite{Jinnouchi2019a,Jinnouchi2019b} demonstrated an on-the-fly force-field generation scheme.
The use of GPR for the direct representation of the PES was reported in Refs.~\cite{Bartok2010,Bartok2015,mones2016,Cui2016,Kolb2017,Carrington2023}.
Furthermore, it was employed to accelerate certain computational steps by, for example, evaluating PES matrix elements in a 
convenient format~\cite{GPR-tew} or by accelerating time-dependent dynamics~\cite{Richings2018}. 
For a recent review on the use of GPR in the context of computational chemistry and material research we refer to Ref.~\cite{deringer2021}.

Recently, a combination of the ADGA and GPR method was presented in Ref.~\cite{schmitz2020}. The new approach, dubbed GPR-ADGA, 
employed statistical uncertainties from GPR along-side with averaged vibrational densities calculated with the ADGA as criteria for 
choosing whether SPs should be predicted with inexpensive GPR or calculated with a more accurate yet costly 
electronic structure method. The performance of GPR-ADGA was assessed by computing fundamental excitation energies from generated PESs. It was demonstrated that
GPR-ADGA could reduce the number of SPs by 65\%--90\% while introducing a root mean square deviation (RMSD) in fundamental frequencies below 2~cm$^{-1}$
compared to the standard ADGA. The algorithm, however, was applicable only to about 3 to 4 atoms.
In the current work, we lift this limitation and demonstrate an improved and extended version of GPR-ADGA, which is applied for PES computations of up to 
10-atomic molecules while maintaining a high accuracy in fundamental excitation energies and a large reduction in the number of SPs (when compared to the standard ADGA).

This work is organized as follows. The underlying theory of the GPR-ADGA method together with its recent technical and 
methodological extensions is described in Sec.~\ref{sec:theory}.
The computational details are provided in Sec.~\ref{sec:comp_details} and followed by GPR-ADGA computations of PESs presented in Sec.~\ref{sec:results}.
Subsequently, conclusions to this work are given in Sec.~\ref{sec:concl}.

\section{Theory \label{sec:theory}}

In the following, we briefly summarize GPR-ADGA components such as the $n$-mode expansion~\cite{jung1996,carter1997,bowman2003,rauhut2004,kongsted2006} 
in Sec.~\ref{sec:n_mode_exp}, 
the theory behind the ADGA~\cite{sparta2009,toffoli2011,klinting2018} in Sec.~\ref{sec:theory_adga}, 
and GPR~\cite{rasmussen2005} in Sec.~\ref{sec:theory_gpr}. 
Then in Sec.~\ref{sec:theory_combin}, we describe the main idea of the
GPR-ADGA method and focus on its recent methodological extensions, 
which enable calculations of larger molecules, in Sec.~\ref{sec:theory_large_molec}. 

\subsection{$n$-Mode Expansion \label{sec:n_mode_exp}}

As was described above in Sec.~\ref{sec:intro}, constructions of full-dimensional PESs $V(\mathbf{q})$, depending on $M = 3K_{\mathrm{nuc}}-6(5)$ number of normal 
vibrational coordinates $\mathbf{q} = \{q_1, q_2, \dots, q_M\}$,
are prohibitively expensive for more than about 4 atoms, i.e., for $K_{\mathrm{nuc}} \gtrsim 4$. 
In order to lift this limitation and construct PESs for larger molecular systems, additional approximations need to be invoked. 
To that end, we first define mode combinations (MCs) $\mathbf{m}_k$ as sets $\{m_1, m_2, \dots, m_k \}$ containing $k$ coordinate indices. Subsequently,
mode combination ranges (MCRs) are formed as sets of MCs (for more details on MCs and MCRs, 
see Ref.~\cite{koenig2016}). A full-dimensional PESs $V(\mathbf{q})$ can then be represented as~\cite{koenig2016}, 
\begin{equation}\label{eq:n_mode_exp}
   V(\mathbf{q}) = \sum_{ \mathbf{m}_k \in \mathrm{MCR} }   \bar{V}^{ \mathbf{m}_k  }     
        = \sum_{ \mathbf{m}_k \in \mathrm{MCR}  } \sum_{ \substack{
        \mathrm{\bf m}_{k^{\prime}} \subseteq  \mathrm{\bf m}_k   \\
        \mathrm{\bf m}_{k^{\prime}} \in \mathrm{MCR}  }}  (-1)^{k-k^{\prime}}  V^{ \mathbf{m}_{k^{\prime}}   },
\end{equation}
where the outer sum in the right-hand side runs over all MCs $\mathbf{m}_k$ from the MCR and the inner sum runs over   
all subsets of $\mathbf{m}_k$ (including $\mathbf{m}_k$ itself). In this formulation, the potential $V(\mathbf{q})$ is conveniently represented as 
a sum of its lower-dimensional cuts excluding overcounting of equivalent terms. 
In Eq.~(\ref{eq:n_mode_exp}), the equality sign holds if the MCR contains MCs $\mathbf{m}_k$ of up to $M$th order.
Constructing the MCR from MCs with at most $n$ mode indices (where $n<M$), one 
neglects mode--mode couplings of higher orders and provides an approximate treatment of the PES. 
This approach drastically reduces the number of SPs to be computed (compared to a fully-coupled PES) and 
is often referred to as the $n$-mode approximation~\cite{jung1996,carter1997,bowman2003,rauhut2004,kongsted2006}.
The number of SPs to be computed, when constructing a PES on a grid of points within the 
$n$-mode representation, is given by
\begin{equation}
  N_{\rm{SPs}} = \sum_{k=1}^{n} {M\choose k}  (g_k)^k\quad,
  \label{eq:number_of_sps}
\end{equation}
where $g_k$ is the number of SPs in the direct product grid per MC $\mathbf{m}_k$.
The number $g_k$ required for accurate PES representation is usually unknown and could vary for different regions of the same PES.  
Another complication lies in the constructions of such grids of points. 
Static grids with predefined and equidistantly separated points offer an easy solution to this issue. 
However, this approach is by no means optimal and often results in a large number of SPs to be computed (for example, see Ref.~\cite{arti2020}).

\subsection{Adaptive Density-Guided Approach \label{sec:theory_adga}}

The ADGA~\cite{sparta2009,toffoli2011,klinting2018} is designed for a fully automatic grid construction and has an advantage over above-mentioned static grid approaches.
It calculates PESs employing the $n$-mode representation and an iterative procedure, which is guided by one-mode averaged vibrational densities of the form,
\begin{equation} \label{eq:adga_dens}
    \rho^{\mathrm{ave}}_{\mathrm{iter}} (q_{m_k}) = \frac{1}{N^{m_k}_{\mathrm{modal}}} \sum^{N^{m_k}_{\mathrm{modal}}}_{s^{m_k} =1} 
    | \varphi_{s^{m_k}}^{m_k}  ( q_{m_k}  ) |^2 ,
\end{equation}
which are obtained in each iteration iter from vibrational self-consistent field (VSCF) calculations~\cite{bowman1978,gerber1979,christiansen2004,hansen2010}.
In Eq.~(\ref{eq:adga_dens}), $\varphi_{s^{m_k}}^{m_k}  ( q_{m_k}  )$ are orthonormal one-mode wave functions (modals) used to describe 
a vibrational mode $q_{m_k}$ and $N^{m_k}_{\mathrm{modal}}$ 
is the number of these modals. Averaged vibrational densities $\rho^{\mathrm{ave}}_{\mathrm{iter}} (q_{m_k})$ 
and the corresponding one-mode potentials $V^{m_k}_{\mathrm{iter}} (q_{m_k})$
are then used to calculate 
an energy-like quantity, which, in the one dimensional case, is given by 
\begin{equation} 
    \Xi_{\mathrm{iter}}^{m_k} = \int_{I_{m_k}} \rho^{\mathrm{ave}}_{\mathrm{iter}} (q_{m_k}) 
    V^{m_k}_{\mathrm{iter}} (q_{m_k}) \mathrm{d} q_{m_k}.
    \label{eq:adga_int}
\end{equation}
$\Xi_{\mathrm{iter}}^{m_k}$ is computed for all intervals 
$I_{m_k}$ defined by 
neighboring SPs of mode $q_{m_k}$.  
If $\Xi_{\mathrm{iter}}^{m_k}$, computed for a particular 
interval $I_{m_k}$, changes significantly between two ADGA iterations, 
the corresponding interval is divided at the middle by inserting a new SP. Note that required SPs are computed with external electronic structure programs. 
The procedure stops when no significant changes in 
$\Xi_{\mathrm{iter}}^{m_k}$ are detected for all modes and intervals. 
After that the ADGA continues analogously for higher mode-couplings, until convergence at the specified MC level is 
achieved. Furthermore, the ADGA automatically extends the grid boundaries 
if a non-negligible 
amount of the average vibrational density $\rho^{\mathrm{ave}}_{\mathrm{iter}} (q_{m_k})$ is detected outside the current grid. 
The convergence of the ADGA is controlled with three criteria, 
$\epsilon_{\textup{rel}}$, $\epsilon_{\textup{abs}}$, and $\epsilon_{\rho}$, where
$\epsilon_{\textup{rel}}$ and $\epsilon_{\textup{abs}}$ assess the relative and absolute change of the integral value $\Xi_{\mathrm{iter}}^{m_k}$, respectively, between 
subsequent iterations and $\epsilon_{\rho}$ checks for the amount of vibrational density outside the grid boundaries. 
For more details on these thresholds, we refer to the original works in Refs.~\cite{sparta2009,klinting2018}.

\subsection{Gaussian Process Regression \label{sec:theory_gpr}}

A further speed-up of the PES construction procedure within the ADGA could be achieved by replacing 
some of costly SP computations with inexpensive GPR predictions. To that end, we introduce vectors
of coordinates $\mathbf{x}_i = (x_{i1}, x_{i2}, \cdots, x_{id})^{\mathrm{T}}$ each describing a particular $i$th molecular conformation. 
These vectors $\mathbf{x}_i$ are often referred to as \textit{input} or \textit{feature vectors}~\cite{murphy2012}. 
In our previous studies in Refs.~\cite{schmitz2018,schmitz2019,schmitz2020}, $\mathbf{x}_i$
were minimal sets of internal coordinates. In the current work, we still apply internal coordinates but additionally standardize them by shifting and 
scaling each feature (for more details, 
see Sec.~S1.1 in the SI). Note, however, that normal coordinates $\mathbf{q}$ are employed in the ADGA
computations of this paper, although extension of the ADGA to other coordinates have been reported~\cite{polysphc}.
The known values of the potential $V(\mathbf{x})$ for a given set of molecular structures $\{\mathbf{x}_i  \}_i^N$ are then collected in a vector 
$\mathbf{v} = (V(\mathbf{x}_1), V(\mathbf{x}_2), \cdots, V(\mathbf{x}_N) )^{\mathrm{T}}$, 
which can be regarded as a vector of \textit{outputs} or \textit{labels}~\cite{murphy2012}.
We assume that the elements of $\mathbf{v}$ have a multivariant Gaussian distribution, i.e.,
\begin{equation} \label{eq:multivar_gauss}
    \mathbf{v} \sim \mathcal{N}({\mathbf{m}}, {\mathbf{K}} + \sigma^2_N {\mathbf{I}}), 
\end{equation}
where $\mathbf{m}$ is the prior mean vector of length $N$, $\sigma^2_N$ is a regularization parameter or noise, and $\mathbf{I}$ is the $N \times N$ identity matrix.
$\mathbf{K}$ is the prior $N \times N$ co-variance matrix with elements $(\mathbf{K})_{ij}$ being equal to the kernel function $k(\mathbf{x}_i, \mathbf{x}_j)$
evaluated for molecular structures $\mathbf{x}_i$ and $\mathbf{x}_j$. In many practical applications of GPR, including our previous studies 
in Refs.~\cite{schmitz2018,schmitz2019,schmitz2020},
the mean vector $\mathbf{m}$ is set to zero. However, in the current work, the components of $\mathbf{m}$
are given as potential energy functions for the quantum mechanical harmonic oscillator,
\begin{equation} \label{eq:prior_mean}
    m(\mathbf{x}_i) = E_0 + \frac{1}{2} (\mathbf{x}_i- \mathbf{x}_0)^{\mathrm{T}} \, \mathbf{H} \, (\mathbf{x}_i - \mathbf{x}_0),
\end{equation}
with $\mathbf{x}_0$ and $E_0$ being the optimized molecular structure in a minimal set of internal coordinates
and the corresponding reference energy, respectively, and $\mathbf{H}$ being the matrix  
of second derivatives of energy at $\mathbf{x}_0$ with respect to molecular displacements (i.e., the Hessian matrix).
Therefore, the GPR-based approach employed in this work predicts deviations of $V(\mathbf{x})$ from a harmonic potential given in Eq.~({\ref{eq:prior_mean}}) and  
could be regarded as a variant of $\Delta$-learning~\cite{ramak2015} or semi-parametric GPR~\cite{rasmussen2005}.  
It is, of course, trivial to extend the procedure to other $m(\mathbf{x}_i)$.

In order to predict unknown values of the potential $\mathbf{v}^* = (V(\mathbf{x}_1^*), V(\mathbf{x}_2^*), \cdots, V(\mathbf{x}_{N^*}^*) )^{\mathrm{T}}$
for a set of $N^*$ molecular structures $\{\mathbf{x}_i^*  \}_i^{N^*}$, 
the joint Gaussian distribution is conditioned on the observations, i.e.,
\begin{equation}
   \mathbf{v}^{*}|{\mathbf{v}} \sim \mathcal{N}({\pmb{\mu}}, {\pmb{\Sigma}}).
\end{equation}
Subsequently, we use the posterior mean vector $\pmb{\mu}$ of length $N^*$ as a predictor~\cite{rasmussen2005},
\begin{equation} \label{eq:posterior_mean}
    \mathbf{v}^{*} \approx {\pmb{\mu}} =  \mathbf{m}^* + (\mathbf{K}^*)^{\mathrm{T}} (\mathbf{K} + \sigma^2_N {\bf{I}})^{-1}  (\mathbf{v} - \mathbf{m}),
\end{equation}
where $\mathbf{K}^*$ denotes the $N \times N^*$ matrix of elements $(\mathbf{K}^*)_{ij} = k(\mathbf{x}_i, \mathbf{x}_j^*)$ and 
$\mathbf{m}^*$ is a vector of length $N^*$ containing mean values 
from Eq.~(\ref{eq:prior_mean}) evaluated for $\mathbf{x}_i^*$.
The diagonal elements $\mathbb{V} [ V(\mathbf{x}_i^*)  ] = ({\pmb{\Sigma}})_{ii}$ of the posterior co-variance matrix~\cite{rasmussen2005},
\begin{equation} \label{eq:posterior_covar}
    {\pmb{\Sigma}} = \mathbf{K}^{**} -  (\mathbf{K}^*)^{\mathrm{T}} (\mathbf{K} + \sigma^2_N {\bf{I}})^{-1} \mathbf{K}^{*},
\end{equation}
are employed as statistical error estimates for the predicted values $V(\mathbf{x}_i^*)$ of the potential. 
Here, $\mathbf{K}^{**}$ is the $N^* \times N^*$ matrix  
of elements
$(\mathbf{K}^{**})_{ij} = k(\mathbf{x}_i^*, \mathbf{x}_j^*)$.

\subsection{Combined GPR-ADGA Methodology \label{sec:theory_combin}}

As was mentioned above, the general idea of GPR-ADGA~\cite{schmitz2020} lies in using the GPR variance 
for a predicted potential value $V(\mathbf{q}_i^*)$
with the corresponding averaged VSCF vibrational density  
as criteria for choosing whether the predicted value $V(\mathbf{q}_i^*)$ should be included in the PES as is or re-calculated with a more accurate and expensive 
electronic structure method. 
In practice, the procedure is carried out in an iterative manner
and starts from training the GPR predictor on the available dataset of coordinates $\mathbf{q}_i$ and corresponding energy values $V(\mathbf{q}_i)$ 
(and possibly energy derivatives with respect to $\mathbf{q}_i$) computed with an electronic structure method.
Then, the ADGA calculation is carried out until its full convergence using GPR-predicted SPs. 
For each SP of the constructed PES, the following quantity, 
\begin{equation} \label{eq:gpr_adga_criterion}
    \Omega^{\mathbf{m}_n}_{\mathrm{box}_i} (\mathbf{q}_i^*) 
    = A^{\mathbf{m}_n}  \rho^{\mathbf{m}_n} 
     \mathbb{V} [ V(\mathbf{q}_i^*)  ], 
\end{equation}
is evaluated. In Eq.~(\ref{eq:gpr_adga_criterion}),  
$A^{\mathbf{m}_n}$ is the box size (i.e., length, area or volume for one-, two- or three-dimensional potential cuts, 
respectively) equal to $\int_{\mathrm{box}_i} \mathrm{d} q_{m_1}, \mathrm{d} q_{m_2}, \dots, \mathrm{d} q_{m_n}$, 
$\rho^{\mathbf{m}_n}$ 
is a vibrational density computed as 
a product $\prod_{m_k \in \mathbf{m_n}} \rho^{\mathrm{ave}}_{\mathrm{iter}} (q_{m_k})$ 
of one-mode VSCF averaged vibrational densities $\rho^{\mathrm{ave}}_{\mathrm{iter}} (q_{m_k})$, and  
$\mathbb{V} [ V(\mathbf{q}_i^*) ]$ is the GPR variance for the predicted energy value  
$V(\mathbf{q}_i^*)$. SPs, for which the value of $\Omega^{\mathbf{m}_n}_{\mathrm{box}_i} (\mathbf{q}_i^*)$ is larger than the specified threshold $T_{\Omega}$
(and some other selection rules are fulfilled; see Ref.~\cite{schmitz2020}),
are collected in a list. All SPs from this list are then re-calculated with electronic structure method and added to the existing dataset of energy values.
Subsequently, the steps including the GPR predictor training, ADGA computation, and another selection of SPs are repeated. 
The procedure is continued until the list of SPs, for which $\Omega^{\mathbf{m}_n}_{\mathrm{box}_i} (\mathbf{q}_i^*) > T_{\Omega}$, is empty. 

\subsection{Extension to Larger Molecules \label{sec:theory_large_molec}}

A large number of technical and methodologocal modifications to original GPR-ADGA~\cite{schmitz2020} in the 
Molecular Interactions Dynamics And Simulation Chemistry Program Package 
({\sc MidasCpp})~\cite{midascpp} was done within this work 
to enable computations of PESs for up to 10 atoms (for examples of PES computations, see Sec.~\ref{sec:results}). 
The former include 
i) predicting energy values $V(\mathbf{q}_i^*)$ in batches for better stability and easier parallelization, 
ii) OpenMP parallelization over the number of batches,
iii) more efficient use of memory and disk space, 
iv) implementation of initial guesses for hyperparameter optimization (see Sec.~S1.2 in the SI), 
as well as a general clean-up removing redundant steps and improving the overall performance. 
Methodological changes require a more detailed consideration.

As was mentioned already in Sec.~\ref{sec:theory_gpr} and shown in Eq.~(\ref{eq:prior_mean}), we use a variant of $\Delta$-learning
technique by predicting the difference between the actual PES and a PES described within the harmonic approximation.
This choice has two advantages. Thus, by predicting the anharmonicity correction instead of the full PES, we potentially also decrease errors in predicted values. 
Furthermore, this allows us using the derivative information for the reference point $\mathbf{x}_0$ 
in the GPR mean function as opposed to placing it directly into the training set (as was done in our previous works in Refs.~\cite{schmitz2019,schmitz2020}).  
Therefore, we avoid enlarging the training set size.

Another important methodological change is related to hyperparameters optimization. In our previous works in Refs.~\cite{schmitz2019,schmitz2020}, 
we used kernels with one signal variance $\sigma_f^2$ and $M$-number of characteristic length-scale parameters $l_i$, each being optimized for 
an individual degree of freedom. This provided a better optimization flexibility and allowed GPR to 
adjust to the physical nature of each coordinate. However, with the number of atoms growing, hyperparameter optimization quickly becomes 
prohibitively expensive, whereas a large number of parameters to be optimized leads to numerical instabilities and multiple minimas being 
present on the optimized hypersurface. These limitation were lifted by standardizing coordinates, i.e., by shifting and scaling them such that
a single length-scale parameter $l$ could be applied for all degrees of freedom (for more details on data standardization, see Sec.~S1.1 in the SI).
This enabled the use of simpler kernels, such as the squared exponential kernel,
\begin{equation}
    k(\mathbf{x}_i, \mathbf{x}_j) = \sigma_f^2 \exp \left( - \frac{(\mathbf{x}_i - \mathbf{x}_j )^2}{2l^2}   \right), 
\end{equation}
which was employed in this work. 
Furthermore, we implemented an additional criterion controlling the hyperparameter optimization procedure. In our setup, hyperparameter optimization is not performed 
during a GPR iteration if a pre-defined number of SPs per 2M-cut is present in the training dataset, i.e., hyperparameters are only optimized for the first few GPR iterations 
and kept fixed afterwards. 
This allows for yet another reduction in the overall computational cost since the hyperparameter optimization is more demanding for larger training set sizes.

\section{Computational Details \label{sec:comp_details}}

\begin{figure}[!h]
\centering
\includegraphics[width=1.0\textwidth]{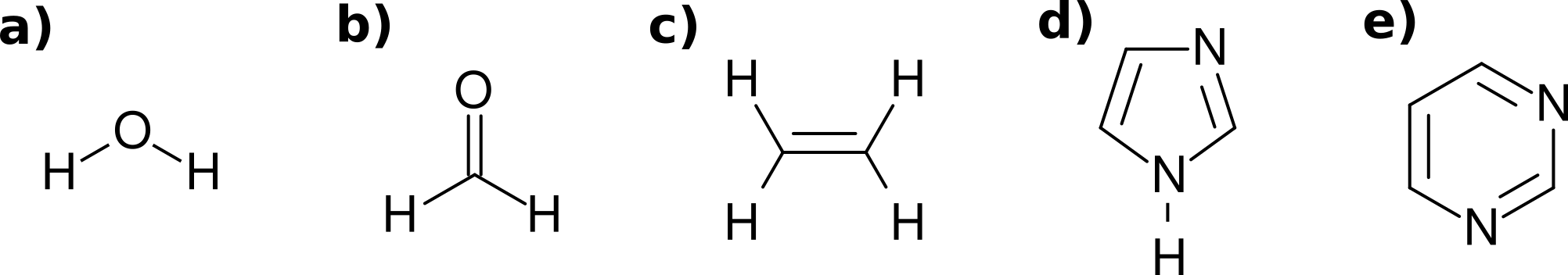}
\caption{Lewis structures of a) water, b) formaldehyde, c) ethylene, d) imidazole, and e) pyrimidine used in this work.}
\label{fig:mol_str}
\end{figure}

To validate the new variant of GPR-ADGA, we chose a test set of molecules of growing size (from 3 to 10 atoms). This set includes 
water, formaldehyde, ethylene, imidazole, and pyrimidine (for Lewis structures, see Fig.~\ref{fig:mol_str}).
The corresponding molecular structures were optimized in {\sc Orca}~\cite{orca} using the Hartree--Fock with three corrections (HF-3c) approach~\cite{sure2013}.
Subsequently, same electronic structure method was applied to compute SPs and second derivatives of energy as presented in Sec.~\ref{sec:5_molec}.

Generation of normal coordinates and construction of PESs with the ADGA~\cite{sparta2009,toffoli2011,klinting2018} 
and GPR-ADGA~\cite{schmitz2020} were carried out in the  
{\sc MidasCpp}~\cite{midascpp}. 
In all presented PES computations, the $n$-mode expansion was truncated at the second order.
To find the optimal set of ADGA thresholds~\cite{sparta2009,klinting2018}, in terms of accuracy of fundamental excitations and the computational cost, 
we carried out a small benchmark study using the above-mentioned test set of molecules and ten sets of ADGA criteria. 
The raw data from this benchmark is given in Sec.~S2 in the SI. The use of thresholds 
$\epsilon_{\textup{rel}} = 1.0 \times 10^{-2}$, 
$\epsilon_{\textup{abs}} = 1.0 \times 10^{-5}$, and
$\epsilon_{\rho} = 1.0 \times 10^{-3}$ 
led to minimum numbers of SPs being computed, while producing  
RMSDs in fundamental excitations below 1~cm$^{-1}$.
For this reason, these thresholds 
were selected as optimal and applied in all GPR-ADGA and reference ADGA computations (unless stated otherwise). 
For the determination of initial grid boundaries, harmonic oscillator turning points with the quantum number $v=2$ were applied.
Four VSCF modals were included in the mean density given in Eq.~(\ref{eq:adga_dens}).
Reference ADGA computations were carried out using the dynamic extension of grid boundaries and 
gradient-guided basis set determination as described in Ref.~\cite{klinting2018}. 
Same settings were found to be incompatible with GPR-ADGA and were disabled.
Contrary to reference ADGA computations, GPR-ADGA was allowed to extend the potential grid boundaries already from the first iteration (by default, 
ADGA does so starting from iteration two and onwards).

To provide the initial training set of points for GPR during the first GPR-ADGA iteration,
the static grid approach~\cite{toffoli2007} was used. The constructed initial static grids contained two SPs per each one-mode combination and the reference 
SP energy (i.e., 7, 13, 25, 43, and 49 SPs for water, formaldehyde, ethylene, imidazole, and pyrimidine, respectively).  
The corresponding static grid boundaries were set using the harmonic oscillator turning points defined by quantum number $v=10$. 
Second derivatives of energy with respect to molecular displacements (Hessian) at the 
equilibrium molecular structure were computed and used to set up the mean function as described in Sec.~\ref{sec:theory_gpr}. 
Furthermore, one-mode grid boundaries were added to the GPR-ADGA list of points, 
which are to be calculated, during the first GPR iteration. The training data, both features and labels, was standardized 
shifting by mean values and scaling by population standard deviations (for more details on standardization options, see Sec.~S1.1 in the SI).
Shifting and scaling factors were computed using the whole training set of data (no special treatment of outliers).
Note that calculations of new standardization factors and re-standardization of data were carried each time the training set was extended.
Similar to Ref.~\cite{schmitz2020}, the Bunch--Kaufmann decomposition was applied to solve the GPR linear system of equations.
Due to several methodological changes to GPR-ADGA, old thresholds such as values for the selection criterion $T_{\Omega}$ and the noise term $\sigma_N^2$,
found to be optimal in Ref.~\cite{schmitz2020}, were not applicable to the current setup. In order to find new optimal values of $T_{\Omega}$ and $\sigma_N^2$,
we carried out GPR-ADGA computations for the water molecule varying the both criteria independently 
by a factor of ten from $1.0 \times 10^{-7}$ to $1.0 \times 10^{-14}$ (see Sec.~S3 in the SI). 
The best trade-off between accuracy and performance was obtained when $T_{\Omega}$ was equal to $\sigma_N^2$.
Furthermore, the both criteria being simultaneously varied from 
$1.0 \times 10^{-8}$ to $1.0 \times 10^{-11}$ provided a series of computations with growing computational 
cost and accuracy and consistently converging to the reference ADGA. 
In the following, series of such GPR-ADGA computations are demonstrated and discussed in Sec.~\ref{sec:results}.
The hyperparameters were optimized by minimazing the negative logarithm of the marginal likelihood~\cite{rasmussen2005} 
with the iRprop algorithm~\cite{Igel2000}. To find the optimal number of SPs per two-mode cut, which can be used as the threshold for stopping the hyperparameter optimization 
and subsequently re-using hyperparameters (see Sec.~\ref{sec:theory_large_molec}), we tested several values from 0 (no optimization) to 30. 
The results are demonstrated in Sec.~S4 in the SI. The value of 15 was found optimal and was applied in all presented GPR-ADGA calculations.

To obtain analytical representations of PESs, the linear fit using up to 10th order polynomials was applied.
Note that the polynomial order used for fitting depends on the number of SPs and never exceeds it. 
The polynomial order increases with the number of SPs up to the specified value of 10. For high-mode potentials, an additional
cut-off controls that the combination of polynomials does not exceed the 10th order (for more details, see the SI of Ref.~\cite{klinting2018}). 
The fitted PESs were used to compute fundamental excitation energies with VSCF~\cite{bowman1978,gerber1979,christiansen2004,hansen2010}.
The accuracy of GPR-ADGA was assessed by computing maximal, minimal, and root-mean-square deviations 
in fundamental excitations with respect to the reference ADGA. Additionally, 
kernel density estimation (KDE) curves were constructed for the difference between ADGA and GPR-ADGA fundamental excitations, i.e., for 
$\Delta \omega = \omega(\text{ADGA}) - \omega(\text{GPR-ADGA})$ using the
\textsc{Seaborn}~\cite{waskom2021} library. For this purpose, a Gaussian-type kernel was used.

For a further demonstration of the GPR-ADGA computational cost in Sec.~\ref{sec:comp_cost}, the molecular structure of ethylene was re-optimized in
the \textsc{Turbomole} program package V7.0~\cite{turbo}. For this purpose, the Hartree--Fock (HF) method from the \textsc{dscf} module~\cite{ahlrichs1989}
as well as explicitly correlated versions of the M\o{}ller--Plesset perturbation theory to second order of perturbation
in conjunction with the resolution-of-the-identity approximation~\cite{RI1,RI2,RI3}
(RI-MP2-F12)
and Coupled Cluster with single, double, and perturbative triple 
correction [CCSD(F12$^*$)(T)] from the \textsc{ccsdf12} module~\cite{haettig2000} were employed. 
Note that CCSD(F12$^*$) is also known as CCSD-F12c~\cite{werner2011,werner2012}.
As the basis set, the correlation-consistent polarized valence
double-$\zeta$ cc-pVDZ-F12~\cite{peterson2008} was used in all cases.
In F12 calculations, the complementary auxiliary basis set (CABS) approach~\cite{valeev2004} was adopted. 
The corresponding CABS threshold was set to $1.0\times 10^{-8}$.
Additionally, the frozen core approximation 
excluding all orbitals with energies below $-$3 a.u.\ from the correlation treatment was used.
Same electronic structure methods and settings were applied for subsequent ADGA and GPR-ADGA calculations of PESs. 
Calculations of the prior GPR mean function were carried out using numerical Hessians.
Both ADGA and GPR-ADGA computations were performed in parallel using nodes 
Intel Xeon E5-2680 v2 @ 2.8 GHz/128GB with 20 cores in total.

\section{Results \label{sec:results}}

In the following, we demonstrate the performance of GPR-ADGA for PESs construction using a test set of five 
molecules of growing size and compare it with the standard ADGA in Sec.~\ref{sec:5_molec}. 
For this proof-of-principle study, we carried out inexpensive PES computations employing the 
HF-3c electronic structure method. Further in Sec.~\ref{sec:comp_cost}, we 
analyze the computational cost of GPR-ADGA in detail by performing PES computations for the ethylene molecule using 
a series of different electronic structure methods.

\subsection{Calculations of Potential Energy Surfaces \label{sec:5_molec}}

\begin{figure}[!h]
\centering
\includegraphics[width=1.0\textwidth]{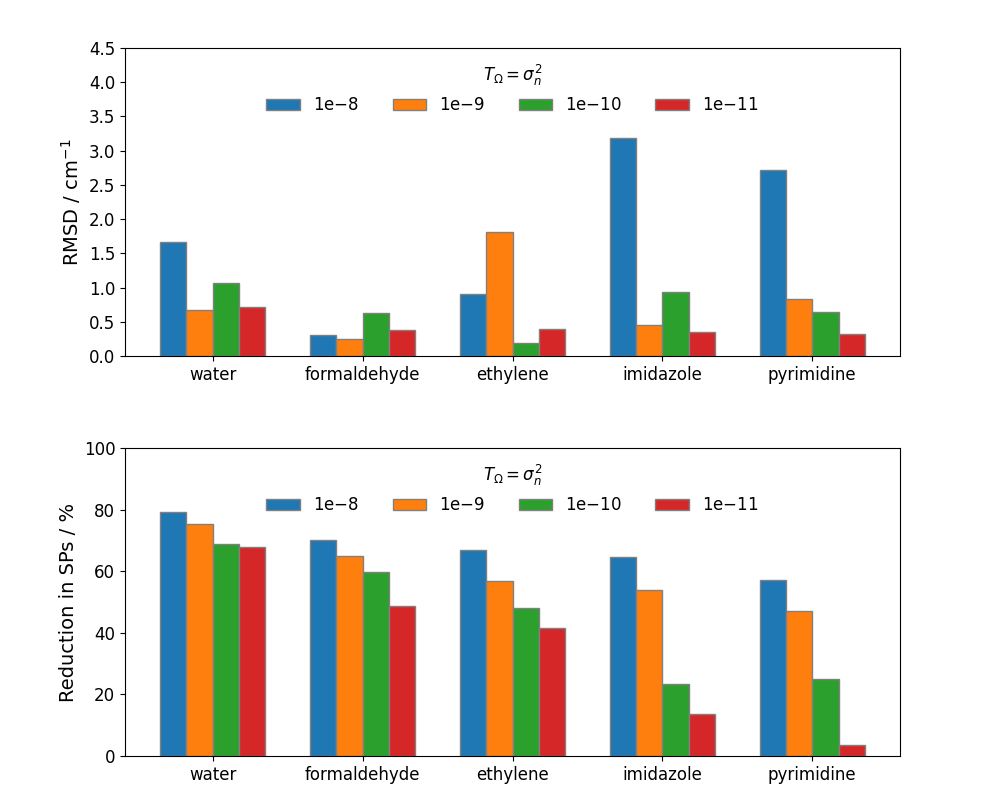}
\caption{Comparison of 2M PESs calculated for the chosen test set of molecules using GPR-ADGA and reference ADGA. 
RMSDs (in cm$^{-1}$) of VSCF fundamental frequencies are shown at the top. Reduction in the number of SPs (in \%) is 
given at the bottom. Results generated with GPR-ADGA criteria $T_{\Omega}$ and $\sigma_N^2$ being simultaneously varied 
in the series  $1.0 \times 10^{-8}$, $1.0 \times 10^{-9}$, $1.0 \times 10^{-10}$, and $1.0 \times 10^{-11}$ are shown in  
blue, orange, green, and red colors, respectively.}
\label{fig:rmsd_sps}
\end{figure}

Performance of GPR-ADGA compared to the reference ADGA is demonstrated in Figs.~\ref{fig:rmsd_sps} (top) and (bottom).
As can be seen and as was expected, the use of tighter thresholds $T_{\Omega} = \sigma_N^2$ results in 
generally smaller deviations in fundamental excitation energies and larger numbers of SPs being computed. Thus, 
with $T_{\Omega} = \sigma_N^2 = 1.0 \times 10^{-8}$ RMSD values are always below 3.2~cm$^{-1}$, whereas the reduction in the number 
of SPs compared to the ADGA is maximal and reaches about 57--79~\% 
(for original values, see Secs.~S2 and S4 in the SI). For the tightest thresholds 
considered here, i.e., for $T_{\Omega} = \sigma_N^2 = 1.0 \times 10^{-11}$, all deviations are below about 0.7~cm$^{-1}$, 
while the reduction in SPs is the smallest and varies from about 4\% to 68~\%. 
Computational savings are always higher for looser thresholds, whereas the dependence found for RMSD values is not always consistent. 
From Fig.~\ref{fig:rmsd_sps} (bottom),
it can also be seen that the gain in terms of the computational cost strongly and consistently depends on the molecular size.
GPR-ADGA performs the best for smaller molecules such as water, where 68--79~\% of SP calculations are avoided.
Unfortunately, a lower reduction of about 4--57~\% is found for pyrimidine.
Note, however, that for this molecule GPR-ADGA still allows to reach the very high accuracy of $\sim$1~cm$^{-1}$, 
while computing about half the number of SPs required for the ADGA (see results for $T_{\Omega} = \sigma_N^2 = 1.0 \times 10^{-9}$).
This behavior of GPR-ADGA is probably related to the reference ADGA computations 
being more efficient for larger molecules. To demonstrate this, we can calculate approximate numbers of SPs 
computed with standard ADGA per 2M-cut potential. Taking the total numbers of SPs in PESs constructed with the reference ADGA
(Sec.~S2 in the SI) and calculating the number of two-mode combinations as $\binom{M}{2} = M!/[2!(M-2)!]$, we can 
verify that about 179, 121, 63, 81, and 58 SPs per 2M-cut function are computed for water,  
formaldehyde, ethylene, imidazole, and pyrimidine, respectively. Therefore, we could conclude that the ADGA
computes smaller numbers of SPs per mode combination of larger molecules, while still reaches the same level of accuracy. 
As the result, a rather modest additional reduction in the number of SPs could be achieved by means of GPR-ADGA.

\begin{figure}[!h]
\centering
\includegraphics[width=1.0\textwidth]{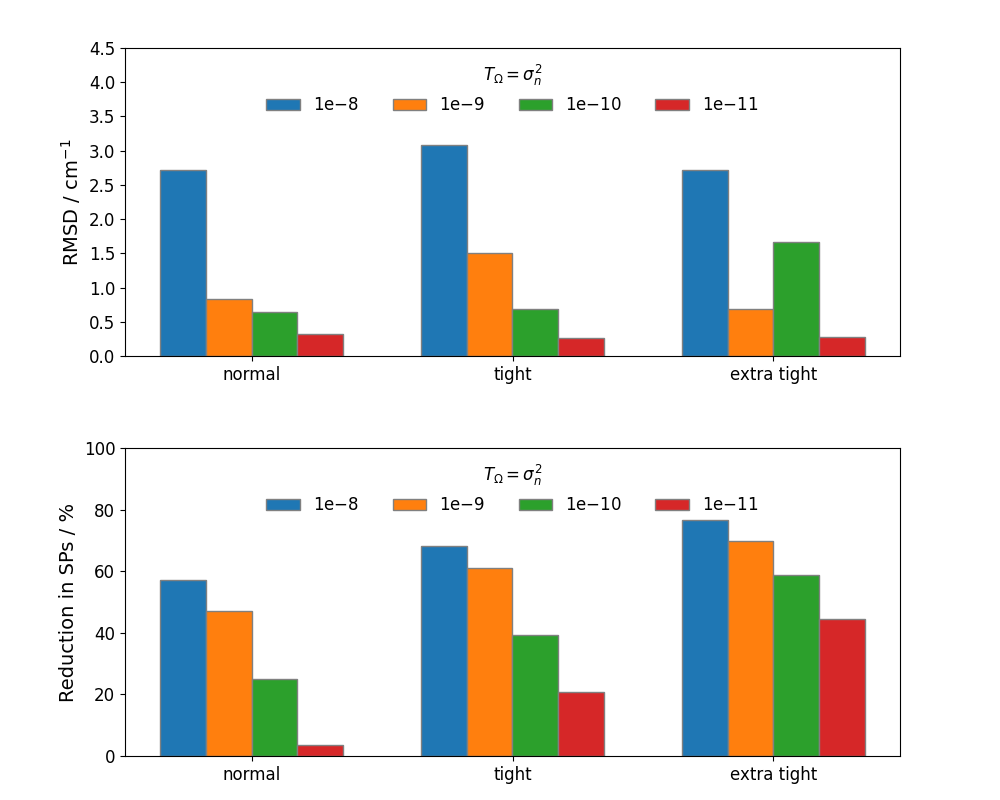}
\caption{Comparison of 2M PESs calculated for pyrimidine using
GPR-ADGA and reference ADGA. RMSDs (in cm$^{-1}$) of VSCF fundamental frequencies are 
shown at the top. Reduction in the number of SPs (in \%) is given at the bottom.
The employed sets of ADGA convergence thresholds (for both, GPR-ADGA and reference ADGA computations) are denoted 
``normal'' ($\epsilon_{\textup{rel}} = 1.0\times 10^{-2}$, $\epsilon_{\textup{abs}} = 1.0\times 10^{-5}$, $\epsilon_{\rho} = 1.0\times 10^{-3}$),
``tight'' ($\epsilon_{\textup{rel}} = 1.0\times 10^{-3}$, $\epsilon_{\textup{abs}} = 1.0\times 10^{-5}$, $\epsilon_{\rho} = 1.0\times 10^{-3}$), 
and ``extra tight'' ($\epsilon_{\textup{rel}} = 1.0\times 10^{-2}$, $\epsilon_{\textup{abs}} = 1.0\times 10^{-6}$, $\epsilon_{\rho} = 1.0\times 10^{-3}$).
Results generated with GPR-ADGA criteria $T_{\Omega}$ and $\sigma_N^2$ being simultaneously varied in the 
series $1.0 \times 10^{-8}$, $1.0 \times 10^{-9}$, $1.0 \times 10^{-10}$, and $1.0 \times 10^{-11}$ are shown in  
blue, orange, green, and red colors, respectively.}
\label{fig:rmsd_sps_vs_adga_crit}
\end{figure}

It is also interesting to compare the performance of the ADGA and GPR-ADGA for the largest molecule from our test set, pyrimidine, while 
varying the convergence thresholds $\epsilon_{\textup{rel}}$, $\epsilon_{\textup{abs}}$, and $\epsilon_{\rho}$. 
Results of this analysis are presented in Fig.~\ref{fig:rmsd_sps_vs_adga_crit}, whereas original values are given in Sec.~S5 in the SI.
As can be seen, RMSD values do not strongly depend on the ADGA convergence criteria
$\epsilon_{\textup{rel}}$, $\epsilon_{\textup{abs}}$, and $\epsilon_{\rho}$
and change by at most $\sim$ 1 cm$^{-1}$. 
Contrary to that, the reduction in the number of SPs changes from 4--57~\% (for ``normal'') to 44--76~\% (for ``extra tight'').
This trend can be explained by the fact that the number of SPs calculated with the reference ADGA is larger for tighter convergence 
thresholds, whereas GPR-ADGA shows a rather weak dependence on $\epsilon_{\textup{rel}}$, $\epsilon_{\textup{abs}}$, and $\epsilon_{\rho}$
(see Sec.~S5 in the SI). For example, the number of SPs computed with the reference ADGA grows by about a factor of two (from 16,022 to 33,836 SPs) when the criteria are changed 
from ``normal'' to ``extra tight''. At the same time, only a rather modest increase of about 18~\% is found for GPR-ADGA using     
$T_{\Omega} = \sigma_N^2 = 1.0\times 10^{-11}$ (from 15,447 to 18,833 SPs).
These results again support the previously discussed point that the relatively smaller gain of GPR-ADGA for large molecules, when compared to the ADGA, is 
related to a higher efficiency of the reference ADGA rather than drawbacks of the GPR-ADGA methodology.

\begin{figure}[!h]
\centering
\includegraphics[width=.8\textwidth]{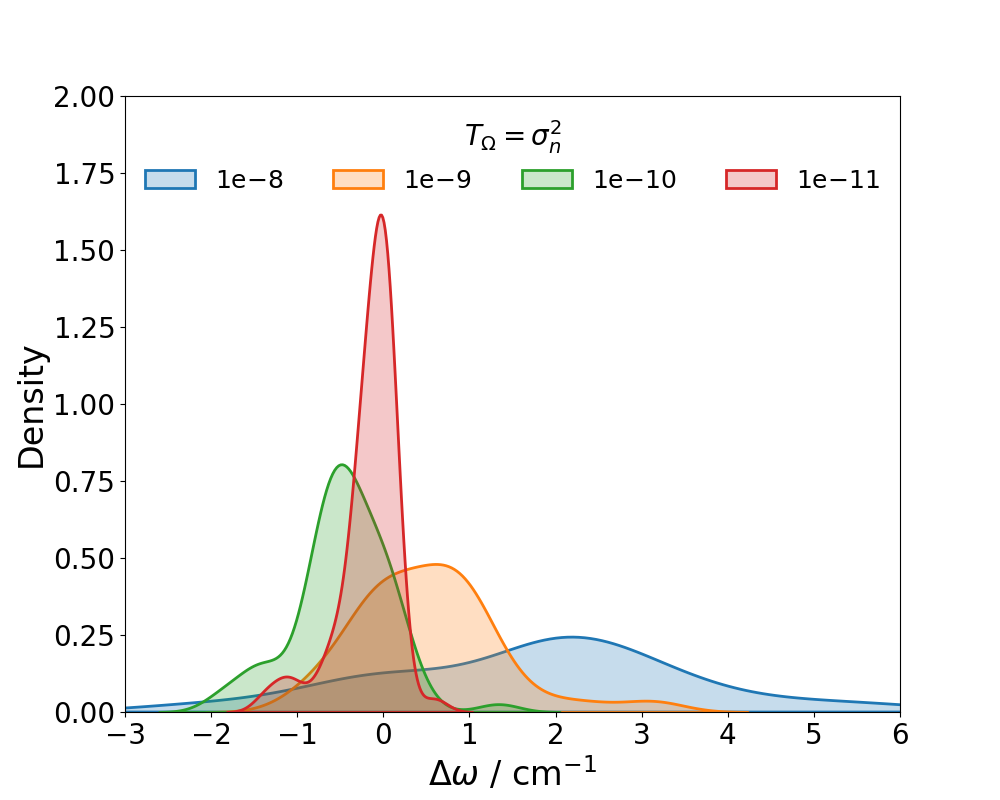}
\caption{KDE curves for deviations in VSCF fundamental frequencies $\Delta \omega = \omega(\text{ADGA}) - \omega(\text{GPR-ADGA})$ (in cm$^{-1}$) 
calculated for all five molecules from the test set. 
Results generated with GPR-ADGA criteria $T_{\Omega}$ and $\sigma_N^2$ being simultaneously varied in the series $1.0 \times 10^{-8}$, 
$1.0 \times 10^{-9}$, $1.0 \times 10^{-10}$, and $1.0 \times 10^{-11}$ are shown in  
blue, orange, green, and red colors, respectively.}
\label{fig:kde}
\end{figure}

To further analyze the accuracy and precision aspects of the GPR-ADGA method, we present KDE curves demonstrating distributions
of errors in VSCF fundamental frequencies in Fig.~\ref{fig:kde}. 
Results are generated for a set of deviations in fundamental excitations belonging to all five molecules from our test set.
In the presented KDE curves, the position of function's maximal value (i.e., the position of the peak) corresponds to the error $\Delta \omega$ with the largest probability.
The closer the position of the peak to the zero at the $\Delta \omega$-axis is, the more accurate results are obtained. 
The precision is reflected in the KDE curve's broadness: A broader 
KDE curve corresponds to a larger error distribution and a lower precision, whereas, on the opposite, a narower curve
indicates at a higher precision. As can be seen from Fig.~\ref{fig:kde}, both the accuracy and precision of GPR-ADGA are consistently
improving for tighter thresholds $T_{\Omega} = \sigma_N^2$.
Thus, a very broad KDE curve with a maximum at about 2.2 cm$^{-1}$ is obtained for $T_{\Omega} = \sigma_N^2 = 1.0 \times 10^{-8}$.
With $T_{\Omega} = \sigma_N^2 = 1.0 \times 10^{-9}$ and $1.0 \times 10^{-9}$, much more narrow KDE curves with the largest probability errors of only about 0.6~cm$^{-1}$ 
and $-0.5$~cm$^{-1}$, respectively, are obtained. Finally, the best results in terms of accuracy and precision are found for $T_{\Omega} = \sigma_N^2 = 1.0 \times 10^{-11}$. 

\subsection{Computational Cost \label{sec:comp_cost}}

As was demonstrated in Sec.~\ref{sec:5_molec}, GPR-ADGA allows to reduce the number of computed SPs by up to 68~\% (compared to the reference ADGA), 
while keeping the RMSD in fundamental excitation energies below 0.7~cm$^{-1}$. If a lower accuracy of about 3~cm$^{-1}$ is considered sufficient,
even larger reduction in the number of SPs of up to about 80~\% could be reached. 
One might argue, however, that a smaller number of SPs being computed does not necessarily mean a faster PES construction procedure as
GPR-ADGA has a much larger computational overhead than the ADGA. This argumentation is correct and the GPR-ADGA method could indeed be more expensive than
the standard ADGA. Thus, GPR-ADGA executes several ADGA computations until their full convergence using GPR as a provider of new SPs. This means, that   
VSCF calculations and fitting of constructed PESs are repeated multiple times on each GPR iteration. 
The number of these iterations could be considerable and reach up to 40 for very tight convergence criteria $T_{\Omega} = \sigma_N^2 = 1.0 \times 10^{-11}$.
Furthermore, additional computational cost is introduced with the use of the GPR algorithm. In this regard, the most expensive steps are the inversion of 
the covariance matrix, i.e., calculations of the term $(\mathbf{K} + \sigma^2_N {\bf{I}})^{-1}$, 
and the solution of an equivalent system of linear equations to find weights 
${\pmb{\omega}} =  (\mathbf{K} + \sigma^2_N {\bf{I}})^{-1}  (\mathbf{v} - \mathbf{m})$ as seen from Eqs.~(\ref{eq:posterior_mean}) and (\ref{eq:posterior_covar}).
Both steps scale as $\mathcal{O}(N^3)$ with the number of training points $N$.
For computing predictions ${\pmb{\mu}}$ and uncertainties $({\pmb{\Sigma}})_{ii}$, it is sufficient to calculate weights ${\pmb{\omega}}$ only once per GPR iteration.
However, the hyperparameter optimization procedure updates hyperparameters and, therefore, requires
re-computing the inverse on each optimization cycle. Since the training set size $N$ grows from iteration to iteration, 
the overhead of using GPR increases as well.
Although in our setup the hyperparameter optimization is not carried out for latter GPR iterations featuring the largest training sets, 
as was described in Sec.~\ref{sec:theory_large_molec}, it still affects the total computational cost of GPR-ADGA.
Finally, the matrix--matrix multiplication $(\mathbf{K} + \sigma^2_N {\bf{I}})^{-1} \mathbf{K}^{*}$ from Eq.~(\ref{eq:posterior_covar}) 
scales as $\mathcal{O}(N^2 N^*)$, where $N^*$ is often much larger than $N$ in practical applications of 
GPR-ADGA. Performing GPR predictions in batches with an OpenMP parallelization over the number of these batches, 
as already mentioned in Sec.~\ref{sec:theory_large_molec}, offers a way to mitigate this step.
Note that the second matrix--matrix multiplication in Eq.~(\ref{eq:posterior_covar})
involving $(\mathbf{K}^{*})^{\mathrm{T}}$ formally scales as $\mathcal{O}(N N^{*2})$. However, because only diagonal elements of 
the posterior covariance matrix ${\pmb{\Sigma}}$ are required, the actual computational scaling is   
reduced to $\mathcal{O}(N N^*)$.

\begin{figure}[!h]
\centering
\includegraphics[width=0.99\textwidth]{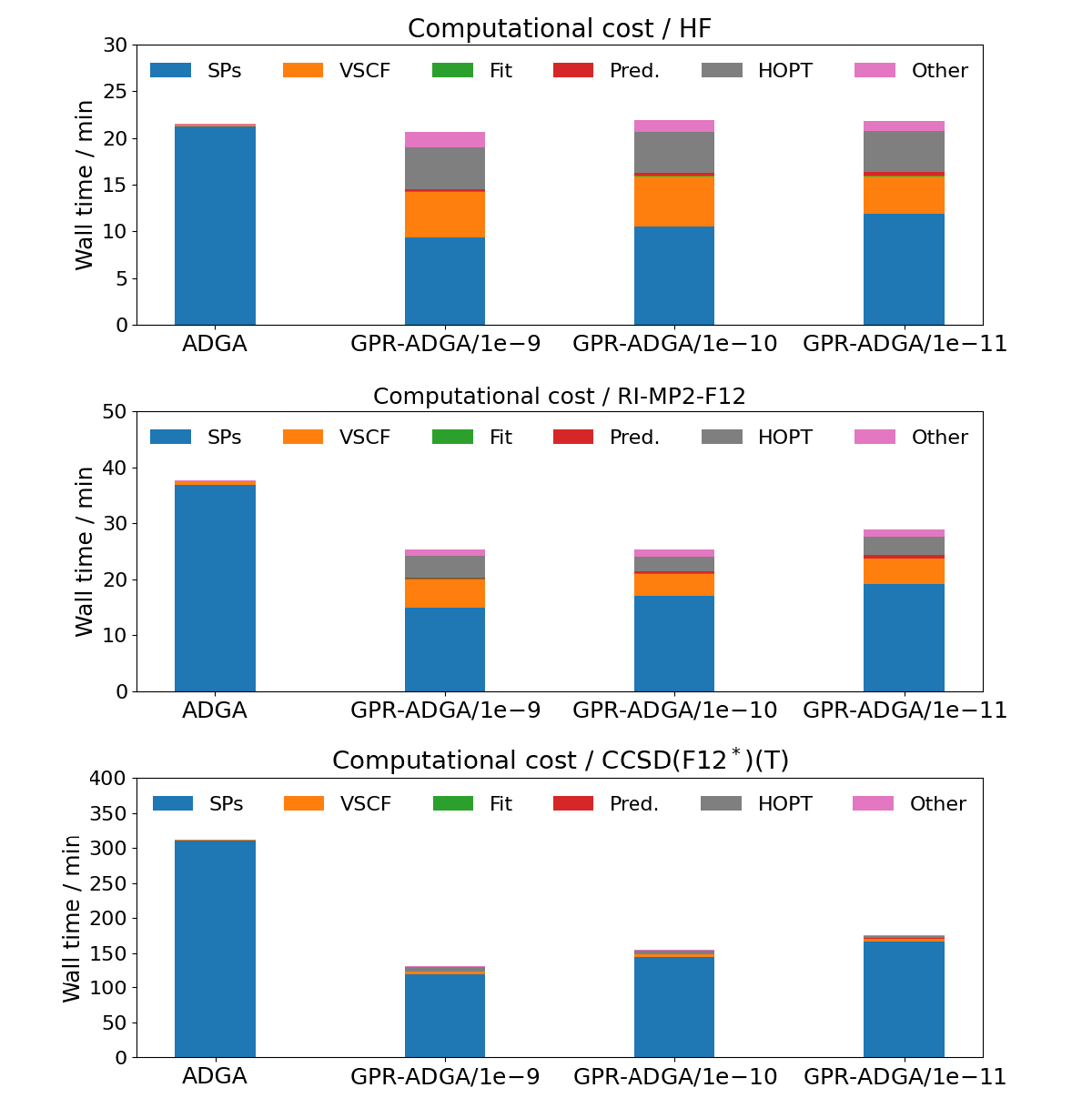}
\caption{Wall times required for GPR-ADGA and reference ADGA computations of ethylene PESs. 
Results are provided for the HF (top), RI-MP2-F12 (middle), and CCSD(F12$^*$)(T) (bottom) methods.
GPR-ADGA criteria $T_{\Omega}$ and $\sigma_N^2$ are simultaneously varied in the series 
$1.0 \times 10^{-9}$, $1.0 \times 10^{-10}$, $1.0 \times 10^{-11}$.
Times spent on SPs, VSCF, polynomial fit of the PES, energy predictions, hyperparameter optimization (denoted as HOPT), and other operations are given in
blue, orange, green, red, grey, and pink colors, respectively.}
\label{fig:comput_cost_ethylene}
\end{figure}

Despite the described above overhead, GPR-ADGA could still considerably reduce the cost of the overall PES construction procedure, 
when expensive electronic structure methods are used. To demonstrate this, we calculated wall times of PESs generation using 
a series of electronic structure methods with increasing computational costs:
HF, RI-MP2-F12, and CCSD(F12$^*$)(T). The results are presented in Fig.~\ref{fig:comput_cost_ethylene}, 
whereas the original values of the wall time as well as CPU time are provided in Sec.~S6 in the SI.
As can be seen from Fig.~\ref{fig:comput_cost_ethylene} (top), with HF being used GPR-ADGA reduces the time spent 
on calculating SPs by about a factor of two (i.e., by $\sim$44--56~\%) compared to the reference ADGA. This, however, does not 
lead to a decreased total computational cost due to a considerable overhead of running multiple 
VSCF calculations and hyperparameter optimizations. These two types of computations amount in 
about 18--24~\% and 20--22~\%, respectively, of the GPR-ADGA wall time. 
As the result, depending on the thresholds being used, GPR-ADGA is comparable or more expensive than the standard ADGA.
The situation changes when RI-MP2-F12 is used for calculating SPs as seen in Fig.~\ref{fig:comput_cost_ethylene} (middle).
The reduction in the SPs computational cost remains about the same, whereas 15--20~\% and 10--15~\% of the GPR-ADGA total computational time is spent
on VSCF and hyperparameter optimization. This leads to GPR-ADGA being 22--32~\% faster than the ADGA. 
Finally, for the most expensive electronic structure method CCSD(F12$^*$)(T), from those applied in this work, 
the cost of SPs becomes dominant in GPR-ADGA with all other computational steps amounting in only about 5--9~\% of the total wall time. 
As the result, GPR-ADGA computation employing CCSD(F12$^*$)(T) is about twice as fast as the reference ADGA using the same electronic structure method. 
Due to a very steep increase in the computational cost of PES construction with the number of atoms, one can expect that the GPR-ADGA overhead becomes 
negligibly small (relative to the cost of SPs) for larger molecules and more expensive electronic structure methods.

In full analogy to the results presented in Sec.~\ref{sec:5_molec}, we also assessed the accuracy of the constructed GPR-ADGA 2M PESs of ethylene 
by computing RMSDs in fundamental excitations and using ADGA as the reference. 
To that end, vibrational coupled cluster with up to two-mode excitations (VCC[2])~\cite{christiansen2004b,christiansen2007,seidler2007,seidler2011,madsen2018b}
was applied. The results are demonstrated in Sec.~S7 in the SI and show very similar trends to those from Sec.~\ref{sec:5_molec}. 
Thus, the RMSD consistently decreases for tighter convergence thresholds $T_{\Omega}$ and $\sigma_N^2$ and reaches values below 1~cm$^{-1}$ for 
$T_{\Omega} = \sigma_N^2 = 1.0 \times 10^{-11}$.

\section{Conclusions \label{sec:concl}}

In this work, we presented a new and improved program implementation of the GPR-ADGA method~\cite{schmitz2020} 
in {\sc MidasCpp}~\cite{midascpp}. A number of technical and methodological extensions was 
introduced enabling GPR-ADGA calculations of PESs for larger molecules than those accessible previously
while maintaining a very high 
accuracy in fundamental excitation energies and a considerable reduction in the number of SPs compared to the reference ADGA.
The performance of GPR-ADGA was assessed on a test set of five molecules of increasing size from three to ten atoms.
Convergence thresholds were introduced allowing one to reach 
a desired balance between the accuracy and the efficiency of the PES construction.
Thus, if the RMSD in fundamental excitation energies of about 
3~cm$^{-1}$ is considered sufficient, calculations of up to 80~\% of SPs could be avoided by using GPR-ADGA.
A higher accuracy of about 0.7~cm$^{-1}$ or better could be reached with tighter GPR-ADGA convergence thresholds
while reducing the number of SPs by up to 68~\%. The reduction in the number of SPs was found to be smaller for larger molecules.
This, however, was explained by a higher efficiency of the reference ADGA for large molecular systems, rather than by   
drawbacks of the GPR-ADGA methodology.

Additionally, we analyzed the computational cost of the PES construction with GPR-ADGA by carrying out calculations of 
ethylene with a series of electronic structure methods such as HF, RI-MP2-F12, and CCSD(F12$^*$)(T) and presenting total wall times.
We showed that due to an increased overhead of GPR-ADGA compared to the reference ADGA
and despite the number of SPs being considerably reduced, 
no computational gain could be reached while using HF. For more accurate and expensive RI-MP2-F12, GPR-ADGA is by 22--32~\% faster than the ADGA 
employing the same electronic structure method. For CCSD(F12$^*$)(T), the overhead of running GPR-ADGA becomes negligibly small compared to 
the wall time spent on SPs. As the result, the computational gain is about the same as the reduction in the number of SPs and reaches 44--58~\%. 
Therefore, we conclude that it is the most advantageous to use GPR-ADGA in conjunction with very accurate and costly electronic structure methods.

Our results demonstrate that GPR-ADGA could be used for highly-accurate and cost-efficient PES calculations
and encourage applications to various molecular systems for subsequent reliable vibrational spectra simulations.
The approach could further be improved by combining it with  
double incremental PES expansions~\cite{koenig2016,arti2020}
and flexible adaptation of local coordinates of nuclei~\cite{koenig2016_falcon}.
This could allow to incorporate fragmentation ideas into GPR-ADGA and handle even larger molecular systems.
In this case, the computational gain could be reached by   
using GPR-ADGA for calculating individual subsystem potentials and/or 
for enabling learning between subsystems of the total molecular system.
The work in both of these directions is currently 
in progress.

\section*{Supplementary Material}
See supplementary material for 
(S1) additional theory aspects,
(S2) reference ADGA-2M computations,
(S3) optimal GPR-ADGA thresholds, 
(S4) parameters controlling hyperparameter optimization,
(S5) influence of ADGA thresholds on GPR-ADGA computations of pyrimidine,
(S6) GPR-ADGA computational cost, and
(S7) vibrational couple cluster computations.

\section*{Acknowledgments}
D.G.A. acknowledges funding from the Rising Star Fellowship of the Department of Biology,
Chemistry, Pharmacy of Freie Universit\"at Berlin.
O.C. acknowledges support from the Independent Research Fund Denmark through grant number 1026-00122B. 
We gratefully acknowledge funding from the Novo Nordisk Foundation
through the Exploratory Interdisciplinary Synergy Programme project
NNF19OC0057790. The authors thank Prof.\ Dr.\ Guntram Rauhut for helpful discussions.

\section*{Data Availability Statement}
The data that support the findings of this study are available from the corresponding author upon reasonable request.

\section*{Author Declarations}
The authors have no conflicts to disclose.

\newpage
\clearpage

%\bibliographystyle{jcp}
%\bibliography{lit}

\begin{thebibliography}{10}

\bibitem{jung1996}
J.~O. Jung, R.~B. Gerber.
\newblock {Vibrational wave functions and spectroscopy of (H$_2$O)$_n$,
  $n$=2,3,4,5: Vibrational self‐consistent field with correlation
  corrections}.
\newblock {\em J. Chem. Phys.}, {\bf 105} (1996) 10332.

\bibitem{carter1997}
S.~Carter, S.~J. Culik, J.~M. Bowman.
\newblock {Vibrational self-consistent field method for many-mode systems: A
  new approach and application to the vibrations of CO adsorbed on Cu(100)}.
\newblock {\em J. Chem. Phys.}, {\bf 107} (1997) 10458.

\bibitem{bowman2003}
J.~M. Bowman, S.~Carter, X.~C. Huang.
\newblock {MULTIMODE: A code to calculate rovibrational energies of polyatomic
  molecules}.
\newblock {\em Int. Rev. Phys. Chem.}, {\bf 22} (2003) 533--549.

\bibitem{rauhut2004}
G.~Rauhut.
\newblock {Efficient calculation of potential energy surfaces for the
  generation of vibrational wave functions}.
\newblock {\em J. Chem. Phys.}, {\bf 121} (2004) 9313--9322.

\bibitem{kongsted2006}
J.~Kongsted, O.~Christiansen.
\newblock {Automatic generation of force fields and property surfaces for use
  in variational vibrational calculations of anharmonic vibrational energies
  and zero-point vibrational averaged properties}.
\newblock {\em J. Chem. Phys.}, {\bf 125} (2006) 124108.

\bibitem{meyer2012}
H.-D. Meyer.
\newblock {Studying molecular quantum dynamics with the multiconfiguration
  time‐dependent Hartree method}.
\newblock {\em Wiley Interdiscip. Rev.: Comput. Mol. Sci.}, {\bf 2} (2012)
  351--374.

\bibitem{rabitz1999}
H.~Rabitz, \"O.~F. Ali\c{s}.
\newblock {General foundations of high‐dimensional model representations}.
\newblock {\em J. Math. Chem.}, {\bf 25} (1999) 197--233.

\bibitem{koenig2016}
C.~K\"onig, O.~Christiansen.
\newblock {Linear-scaling generation of potential energy surfaces using a
  double incremental expansion}.
\newblock {\em J. Chem. Phys.}, {\bf 145} (2016) 064105.

\bibitem{madsen2018}
D.~Madsen, O.~Christiansen, C.~K\"onig.
\newblock Anharmonic vibrational spectra from double incremental potential
  energy and dipole surfaces.
\newblock {\em Phys. Chem. Chem. Phys.}, {\bf 20} (2018) 3445--3456.

\bibitem{yagi2007}
K.~Yagi, S.~Hirata, K.~Hirao.
\newblock {Multiresolution potential energy surfaces for vibrational state
  calculations}.
\newblock {\em Theor. Chem. Acc.}, {\bf 118} (2007) 681--691.

\bibitem{rauhut2008}
G.~Rauhut, T.~Hrenar.
\newblock {A combined variational and perturbational study on the vibrational
  spectrum of P$_2$F$_4$}.
\newblock {\em Chem. Phys.}, {\bf 346} (2008) 160--166.

\bibitem{Rauhut2009}
G.~Rauhut, B.~Hartke.
\newblock {Modeling of high-order many-mode terms in the expansion of
  multidimensional potential energy surfaces: Application to vibrational
  spectra}.
\newblock {\em J. Chem. Phys.}, {\bf 131}(1) (2009) 014108.

\bibitem{sparta2009_2}
M.~Sparta, I.-M. H{\o}yvik, D.~Toffoli, O.~Christiansen.
\newblock {Potential energy surfaces for vibrational structure calculations
  from a multiresolution adaptive density-guided approach: Implementation and
  test Calculations}.
\newblock {\em J. Phys. Chem. A}, {\bf 113} (2009) 8712--8723.

\bibitem{sparta2010}
M.~Sparta, M.~B. Hansen, E.~Matito, D.~Toffoli, O.~Christiansen.
\newblock {Using electronic energy derivative information in automated
  potential energy surface construction for vibrational calculations}.
\newblock {\em J. Chem. Theory Comput.}, {\bf 6} (2010) 3162--3175.

\bibitem{meier2013}
P.~Meier, G.~Bellchambers, J.~Klepp, F.~R. Manby, G.~Rauhut.
\newblock {Modeling of high-order terms in potential energy surface expansions
  using the reference-geometry Harris--Foulkes method}.
\newblock {\em Phys. Chem. Chem. Phys.}, {\bf 15} (2013) 10233--10240.

\bibitem{schmitz2019}
G.~Schmitz, D.~G. Artiukhin, O.~Christiansen.
\newblock {Approximate high mode coupling potentials using Gaussian process
  regression and adaptive density guided sampling}.
\newblock {\em J. Chem. Phys.}, {\bf 150} (2019) 131102.

\bibitem{benoit2004}
D.~M. Benoit.
\newblock {Fast vibrational self-consistent field calculations through a
  reduced mode–mode coupling scheme}.
\newblock {\em J. Chem. Phys.}, {\bf 120} (2004) 562--573.

\bibitem{benoit2006}
D.~M. Benoit.
\newblock {Efficient correlation-corrected vibrational self-consistent field
  computation of OH-stretch frequencies using a low-scaling algorithm}.
\newblock {\em J. Chem. Phys.}, {\bf 125} (2006) 244110.

\bibitem{pele2008}
L.~Pele, R.~B. Gerber.
\newblock {On the number of significant mode--mode anharmonic couplings in
  vibrational calculations: Correlation-corrected vibrational self-consistent
  field treatment of di-, tri-, and tetrapeptides}.
\newblock {\em J. Chem. Phys.}, {\bf 128} (2008) 165105.

\bibitem{benoit2008}
D.~M. Benoit.
\newblock {Fast vibrational calculation of anharmonic OH-stretch frequencies
  for two low-energy noradrenaline conformers}.
\newblock {\em J. Chem. Phys.}, {\bf 129} (2008) 234304.

\bibitem{seidler2009}
P.~Seidler, T.~Kaga, K.~Yagi, O.~Christiansen, K.~Hirao.
\newblock {On the coupling strength in potential energy surfaces for
  vibrational calculations}.
\newblock {\em Chem. Phys. Lett.}, {\bf 483} (2009) 138--142.

\bibitem{cheng2014}
X.~Cheng, R.~P. Steele.
\newblock {Efficient anharmonic vibrational spectroscopy for large molecules
  using local-mode coordinates}.
\newblock {\em J. Chem. Phys.}, {\bf 141} (2014) 104105.

\bibitem{Klinting2020}
E.L. Klinting, O.~Christiansen, C.~K{\"{o}}nig.
\newblock Toward accurate theoretical vibrational spectra: A case study for
  maleimide.
\newblock {\em J. Phys. Chem. A}, 2020) acs.jpca.9b11915.

\bibitem{mackeprang2015}
K.~Mackeprang, V.~H\"anninen, L.~Halonen, H.~G. Kjaergaard.
\newblock {The effect of large amplitude motions on the vibrational intensities
  in hydrogen bonded complexes}.
\newblock {\em J. Chem. Phys.}, {\bf 142} (2015) 094304.

\bibitem{yagi2019}
K.~Yagi, K.~Yamada, C.~Kobayashi, Y.~Sugita.
\newblock {Anharmonic vibrational analysis of biomolecules and solvated
  molecules using hybrid QM/MM computations}.
\newblock {\em J. Chem. Theory Comput.}, {\bf 15} (2019) 1924--1938.

\bibitem{ziegler2018}
B.~Ziegler, G.~Rauhut.
\newblock Rigorous use of symmetry within the construction of multidimensional
  potential energy surfaces.
\newblock {\em The Journal of Chemical Physics}, {\bf 149}(16) (2018) 164110.

\bibitem{sparta2009}
M.~Sparta, D.~Toffoli, O.~Christiansen.
\newblock {An adaptive density-guided approach for the generation of potential
  energy surfaces of polyatomic molecules}.
\newblock {\em Theor. Chem. Acc.}, {\bf 123} (2009) 413--429.

\bibitem{toffoli2011}
D.~Toffoli, M.~Sparta, O.~Christiansen.
\newblock {Accurate Multimode Vibrational Calculations Using A B-spline Basis:
  Theory, Tests and Application to Dioxirane and Diazirinone}.
\newblock {\em Mol. Phys.}, {\bf 109} (2011) 673--685.

\bibitem{klinting2018}
E.~L. Klinting, B.~Thomsen, I.~H. Godtliebsen, O.~Christiansen.
\newblock {Employing general fit-bases for construction of potential energy
  surfaces with an adaptive density-guided approach}.
\newblock {\em J. Chem. Phys.}, {\bf 148} (2018) 064113.

\bibitem{arti2020}
D.~G. Artiukhin, E.~L. Klinting, C.~K\"onig, O.~Christiansen.
\newblock {Adaptive density-guided approach to double incremental potential
  energy surface construction}.
\newblock {\em J. Chem. Phys.}, {\bf 152} (2020) 194105.

\bibitem{Manzhos2006}
S.~Manzhos, T.~Carrington.
\newblock Using neural networks to represent potential surfaces as sums of
  products.
\newblock {\em J. Chem. Phys.}, {\bf 125}(19) (2006) 194105.

\bibitem{Manzhos2008}
S.~Manzhos, T.~Carrington.
\newblock Using neural networks, optimized coordinates, and high-dimensional
  model representations to obtain a vinyl bromide potential surface.
\newblock {\em J. Chem. Phys.}, {\bf 129}(22) (2008) 224104.

\bibitem{Brown2017}
A.~Brown, E.~Pradhan.
\newblock Fitting potential energy surfaces to sum-of-products form with neural
  networks using exponential neurons.
\newblock {\em J. Theor. Comput. Chem.}, {\bf 16}(05) (2017) 1730001.

\bibitem{Pradhan2017}
E.~Pradhan, A.~Brown.
\newblock A ground state potential energy surface for hono based on a neural
  network with exponential fitting functions.
\newblock {\em Phys. Chem. Chem. Phys.}, {\bf 19} (2017) 22272--22281.

\bibitem{BrownHFCO}
E.~Pradhan, A.~Brown.
\newblock Vibrational energies for hfco using a neural network sum of
  exponentials potential energy surface.
\newblock {\em J. Chem. Phys.}, {\bf 144}(17) (2016) 174305.

\bibitem{Pradhan2016}
E.~Pradhan, A.~Brown.
\newblock Neural network exponential fitting of a potential energy surface with
  multiple minima: Application to hfco.
\newblock {\em J. Mol. Spectrosc.}, {\bf 330} (2016) 158--164.

\bibitem{behler2007}
J.~Behler, M.~Parrinello.
\newblock Generalized neural-network representation of high-dimensional
  potential-energy surfaces.
\newblock {\em Phys. Rev. Lett.}, {\bf 98} (2007) 146401.

\bibitem{behler2011}
J.~Behler.
\newblock Neural network potential-energy surfaces in chemistry: a tool for
  large-scale simulations.
\newblock {\em Phys. Chem. Chem. Phys.}, {\bf 13} (2011) 17930--17955.

\bibitem{behler2017}
J.~Behler.
\newblock First principles neural network potentials for reactive simulations
  of large molecular and condensed systems.
\newblock {\em Angew. Chem. Int. Ed.}, {\bf 56}(42) (2017) 12828--12840.

\bibitem{Behler2016}
J.~Behler.
\newblock Perspective: Machine learning potentials for atomistic simulations.
\newblock {\em J. Chem. Phys.}, {\bf 145}(17) (2016) 170901.

\bibitem{Ko2021}
T.~W. Ko, J.~A. Finkler, S.~Goedecker, J.~Behler.
\newblock A fourth-generation high-dimensional neural network potential with
  accurate electrostatics including non-local charge transfer.
\newblock {\em Nature Communications}, {\bf 12}(1) (2021) 398.

\bibitem{rasmussen2005}
C.~E. Rasmussen, C.~K.~I. Williams.
\newblock {\em Gaussian Processes for Machine Learning}.
\newblock Adaptive computation and machine learning. MIT Press, 2005.

\bibitem{Schran2020}
C.~Schran, K.~Brezina, O.~Marsalek.
\newblock Committee neural network potentials control generalization errors and
  enable active learning.
\newblock {\em The Journal of Chemical Physics}, {\bf 153}(10) (2020) 104105.

\bibitem{Jinnouchi2019a}
R.~Jinnouchi, F.~Karsai, G.~Kresse.
\newblock On-the-fly machine learning force field generation: Application to
  melting points.
\newblock {\em Phys. Rev. B}, {\bf 100} (2019) 014105.

\bibitem{Jinnouchi2019b}
R.~Jinnouchi, J.~Lahnsteiner, F.~Karsai, G.~Kresse, M.~Bokdam.
\newblock Phase transitions of hybrid perovskites simulated by machine-learning
  force fields trained on the fly with bayesian inference.
\newblock {\em Phys. Rev. Lett.}, {\bf 122} (2019) 225701.

\bibitem{Bartok2010}
A.~P. Bart\'ok, M.~C. Payne, R.~Kondor, G.~Cs\'anyi.
\newblock Gaussian approximation potentials: The accuracy of quantum mechanics,
  without the electrons.
\newblock {\em Phys. Rev. Lett.}, {\bf 104} (2010) 136403.

\bibitem{Bartok2015}
A.~P. Bart\'ok, G.~Cs\'anyi.
\newblock Gaussian approximation potentials: A brief tutorial introduction.
\newblock {\em Int. J. Quantum Chem.}, {\bf 115}(16) (2015) 1051--1057.

\bibitem{mones2016}
L.~Mones, N.~Bernstein, G.~Cs\'anyi.
\newblock Exploration, sampling, and reconstruction of free energy surfaces
  with gaussian process regression.
\newblock {\em J. Chem. Theory Comput.}, {\bf 12}(10) (2016) 5100--5110.

\bibitem{Cui2016}
J.~Cui, R.~V. Krems.
\newblock Efficient non-parametric fitting of potential energy surfaces for
  polyatomic molecules with gaussian processes.
\newblock {\em J. Phys. B: At., Mol. Opt. Phys.}, {\bf 49}(22) (2016) 224001.

\bibitem{Kolb2017}
B.~Kolb, P.~Marshall, B.~Zhao, B.~Jiang, H.~Guo.
\newblock Representing global reactive potential energy surfaces using gaussian
  processes.
\newblock {\em J. Phys. Chem. A}, {\bf 121}(13) (2017) 2552--2557.

\bibitem{Carrington2023}
T.~Carrington, N.~Yang, S.~Hill, S.~Manzhos.
\newblock {A local Gaussian Processes method for fitting potential surfaces
  that obviates the need to invert large matrices}.
\newblock {\em ChemRxiv}, 2023.

\bibitem{GPR-tew}
J.~P. Alborzpour, D.~P. Tew, S.~Habershon.
\newblock Efficient and accurate evaluation of potential energy matrix elements
  for quantum dynamics using gaussian process regression.
\newblock {\em J. Chem. Phys.}, {\bf 145}(17) (2016) 174112.

\bibitem{Richings2018}
G.~W. Richings, S.~Habershon.
\newblock Mctdh on-the-fly: Efficient grid-based quantum dynamics without
  pre-computed potential energy surfaces.
\newblock {\em J. Chem. Phys.}, {\bf 148}(13) (2018) 134116.

\bibitem{deringer2021}
V.~L. Deringer, A.~P. Bart\'ok, N.~Bernstein, D.~M. Wilkins, M.~Ceriotti,
  G.~Cs\'anyi.
\newblock Gaussian process regression for materials and molecules.
\newblock {\em Chem. Rev.}, {\bf 121}(16) (2021) 10073--10141.

\bibitem{schmitz2020}
G.~Schmitz, E.~L. Klinting, O.~Christiansen.
\newblock {A Gaussian process regression adaptive density guided approach for
  potential energy surface construction}.
\newblock {\em J. Chem. Phys.}, {\bf 153} (2020) 064105.

\bibitem{bowman1978}
J.~M. Bowman.
\newblock {Self‐consistent field energies and wavefunctions for coupled
  oscillators}.
\newblock {\em J. Chem. Phys.}, {\bf 68} (1978) 608--610.

\bibitem{gerber1979}
R.~B. Gerber, M.~A. Ratner.
\newblock {A semiclassical self-consistent field (SC SCF) approximation for
  eigenvalues of coupled-vibration systems}.
\newblock {\em Chem. Phys. Lett.}, {\bf 68} (1979) 195--198.

\bibitem{christiansen2004}
O.~Christiansen.
\newblock {A second quantization formulation of multimode dynamics}.
\newblock {\em J. Chem. Phys.}, {\bf 120} (2004) 2140--2148.

\bibitem{hansen2010}
M.~B. Hansen, M.~Sparta, P.~Seidler, D.~Toffoli, O.~Christiansen.
\newblock {New formulation and implementation of vibrational self-consistent
  field theory}.
\newblock {\em J. Chem. Theory Comput.}, {\bf 6} (2010) 235--248.

\bibitem{murphy2012}
K.~P. Murphy.
\newblock {\em {Machine learning: A probabilistic perspective}}.
\newblock MIT press, 2012.

\bibitem{schmitz2018}
G.~Schmitz, O.~Christiansen.
\newblock {Gaussian process regression to accelerate geometry optimizations
  relying on numerical differentiation}.
\newblock {\em J. Chem. Phys.}, {\bf 148} (2018) 241704.

\bibitem{polysphc}
E.~L. Klinting, D.~Lauvergnat, O.~Christiansen.
\newblock Vibrational coupled cluster computations in polyspherical coordinates
  with the exact analytical kinetic energy operator.
\newblock {\em Journal of Chemical Theory and Computation}, {\bf 16}(7) (2020)
  4505--4520.

\bibitem{ramak2015}
R.~Ramakrishnan, P.~O. Dral, M.~Rupp, O.~A. von Lilienfeld.
\newblock {Big Data Meets Quantum Chemistry Approximations: The
  $\Delta$-Machine Learning Approach}.
\newblock {\em J. Chem. Theory Comput.}, {\bf 11}(5) (2015) 2087--2096.

\bibitem{midascpp}
O.~Christiansen, D.~G. Artiukhin, I.~H. Godtliebsen, E.~M. Gras,
  W.~Gy{\H{o}}rffy, M.~B. Hansen, M.~B. Hansen, E.~L. Klinting, J.~Kongsted,
  C.~K{\"o}nig, D.~Madsen, N.~K. Madsen, K.~Monrad, G.~Schmitz, P.~Seidler,
  K.~Sneskov, M.~Sparta, B.~Thomsen, D.~Toffoli, A.~Zoccante.
\newblock {MidasCpp}, version 2022.10.0.
\newblock https://midascpp.gitlab.io/.

\bibitem{orca}
F.~Neese.
\newblock {The ORCA program system}.
\newblock {\em Comput. Mol. Sci.}, {\bf 2} (2012) 73--78.

\bibitem{sure2013}
R.~Sure, S.~Grimme.
\newblock {Corrected small basis set Hartree‐-Fock method for large systems}.
\newblock {\em J. Comput. Chem.}, {\bf 34} (2013) 1672--1685.

\bibitem{toffoli2007}
D.~Toffoli, J.~Kongsted, O.~Christiansen.
\newblock {Automatic generation of potential energy and property surfaces of
  polyatomic molecules in normal coordinates}.
\newblock {\em J. Chem. Phys.}, {\bf 127} (2007) 204106.

\bibitem{Igel2000}
C.~Igel, M.~H\"{u}sken.
\newblock Improving the rprop learning algorithm.
\newblock In {\em Proceedings of the Second International Symposium on Neural
  Computation}, p. 115--121, 2000.

\bibitem{waskom2021}
M.~L. Waskom.
\newblock {Seaborn: statistical data visualization}.
\newblock {\em J. Open Source Softw.}, {\bf 6}(60) (2021) 3021.

\bibitem{turbo}
{TURBOMOLE V7.0 2015}, a development of {University of Karlsruhe} and
  {Forschungszentrum Karlsruhe GmbH}, 1989--2007, {TURBOMOLE GmbH}, since 2007;
  available from {\tt http://www.turbomole.com}.

\bibitem{ahlrichs1989}
R.~Ahlrichs, M.~B\"ar, M.~H\"aser, H.~Horn, C.~K\"olmel.
\newblock {Electronic structure calculations on workstation computers: The
  program system turbomole}.
\newblock {\em Chem. Phys. Lett.}, {\bf 162}(3) (1989) 165--169.

\bibitem{RI1}
J.~L. Whitten.
\newblock {Coulombic Potential Energy Integrals and Approximations}.
\newblock {\em The Journal of Chemical Physics}, {\bf 58}(10) (1973)
  4496--4501.

\bibitem{RI2}
B.~I. Dunlap, J.~W.~D. Connolly, J.~R. Sabin.
\newblock {On Some Approximations in Applications of X Alpha Theory}.
\newblock {\em The Journal of Chemical Physics}, {\bf 71}(8) (1979) 3396--3402.

\bibitem{RI3}
O.~Vahtras, J.~Alml\"{o}f, M.~W. Feyereisen.
\newblock {Integral Approximations for LCAO-SCF Calculations}.
\newblock {\em Chem. Phys. Lett.}, {\bf 213} (1993) 514--518.

\bibitem{haettig2000}
C.~H\"attig, F.~Weigend.
\newblock {CC2 excitation energy calculations on large molecules using the
  resolution of the identity approximation}.
\newblock {\em J. Chem. Phys.}, {\bf 113} (2000) 5154.

\bibitem{werner2011}
H.-J. Werner, G.~Knizia, F.~R. Manby.
\newblock Explicitly correlated coupled cluster methods with pair-specific
  geminals.
\newblock {\em Molecular Physics}, {\bf 109}(3) (2011) 407--417.

\bibitem{werner2012}
H.-J. Werner, P.~J. Knowles, G.~Knizia, F.~R. Manby, M.~Sch\"utz.
\newblock Molpro: a general-purpose quantum chemistry program package.
\newblock {\em WIREs Comput. Mol. Sci.}, {\bf 2} (2012) 242--253.

\bibitem{peterson2008}
K.~A. Peterson, T.~B. Adler, H.-J. Werner.
\newblock {Systematically Convergend Basis Sets for Explicitly Correlated
  Wavefunctions: The Atoms H, He, B--Ne, and Al--Ar}.
\newblock {\em J. Chem. Phys.}, {\bf 128}(8) (2008) 084102.

\bibitem{valeev2004}
E.~F. Valeev.
\newblock {Improving on the resolution of the identity in linear R12 ab initio
  theories}.
\newblock {\em Chem. Phys. Lett.}, {\bf 395} (2004) 190--195.

\bibitem{christiansen2004b}
O.~Christiansen.
\newblock {Vibrational coupled cluster theory}.
\newblock {\em J. Chem. Phys.}, {\bf 120} (2004) 2149.

\bibitem{christiansen2007}
O.~Christiansen.
\newblock {Vibrational structure theory: new vibrational wave function methods
  for calculation of anharmonic vibrational energies and vibrational
  contributions to molecular properties}.
\newblock {\em Phys. Chem. Chem. Phys.}, {\bf 9} (2007) 2942--2953.

\bibitem{seidler2007}
P.~Seidler, O.~Christiansen.
\newblock {Vibrational excitation energies from vibrational coupled cluster
  response theory}.
\newblock {\em J. Chem. Phys.}, {\bf 126} (2007) 204101.

\bibitem{seidler2011}
P.~Seidler, M.~Sparta, O.~Christiansen.
\newblock {Vibrational coupled cluster response theory: A general
  implementation}.
\newblock {\em J. Chem. Phys.}, {\bf 134} (2011) 054119.

\bibitem{madsen2018b}
N.~K. Madsen, I.~H. Godtliebsen, S.~A. Losilla, O.~Christiansen.
\newblock {Tensor-decomposed vibrational coupled-cluster theory: Enabling
  large-scale, highly accurate vibrational-structure calculations}.
\newblock {\em J. Chem. Phys.}, {\bf 148} (2018) 024103.

\bibitem{koenig2016_falcon}
C.~K\"onig, M.~B. Hansen, I.~H. Godtliebsen, O.~Christiansen.
\newblock {FALCON: A method for flexible adaptation of local coordinates of
  nuclei}.
\newblock {\em J. Chem. Phys.}, {\bf 144} (2016) 074108.

\end{thebibliography}

\end{document}